\documentclass[graybox]{svmult}

\usepackage{helvet}         
\usepackage{courier}        
\usepackage{type1cm}        
\usepackage{makeidx}         
\usepackage{graphicx}        
\usepackage{multicol}        
\usepackage[bottom]{footmisc}

\usepackage{amsmath,amssymb}

\begin{document}

\title*{Equivalence groupoid and enhanced\\ group classification  of a class of generalized Kawahara equations}
\titlerunning{Group classification of a class of generalized Kawahara equations}
\author{Olena~Vaneeva, Olena~Magda and Alexander Zhalij}
\institute{Olena~Vaneeva \at Institute of Mathematics of the National Academy of Sciences of Ukraine,\\ 3 Tereshchenkivska Str., 01024 Kyiv,   Ukraine.
\email{vaneeva@imath.kiev.ua}
\and Olena Magda \at Vadym Hetman Kyiv National Economic University,
54/1 Prospect Peremogy,\\
03680 Kyiv,
Ukraine. \email{olena.magda@gmail.com}
\and Alexander Zhalij \at Institute of Mathematics of the National Academy of Sciences of Ukraine,\\ 3 Tereshchenkivska Str., 01024 Kyiv, Ukraine.
\email{zhaliy@imath.kiev.ua}
}
\maketitle

\abstract{Transformation properties of a class of generalized Kawahara equations  with
time-dependent coefficients are studied.
We construct the equivalence groupoid of the class and prove that this class is not normalized
but can be presented as a union of two disjoint normalized subclasses.
Using the obtained results and properly gauging the arbitrary elements of the class, we carry out its complete group classification, which covers  gaps
in the previous works on the subject.}

\section{Introduction}

The equations of Kawahara type are important models appearing in solitary waves theory.
In the usual sense, solitary waves are nonlinear waves of constant form which decay rapidly in their tail regions. The rate of this decay is usually
exponential. However, under critical conditions in dispersive systems (e.g.,  the magneto-acoustic waves in plasmas, the waves with surface tension,
etc.), unexpected rise of weakly nonlocal solitary waves occurs. These waves consist of a central core which is similar to that of classical
solitary waves, but they are accompanied by copropagating oscillatory tails which extend indefinitely far from the core with a nonzero constant
amplitude. In order to describe and clarify the properties of these waves Kawahara introduced generalized nonlinear dispersive equations which have
a form of the KdV equation with an additional fifth order derivative term, namely,
$$
u_t+\alpha u u_x +\beta u_{xxx}+\sigma u_{xxxxx}=0,
$$
where  $\alpha$, $\beta$ and $\sigma$ are nonzero constants~\cite{Kawahara1972}.  This equation was heavily studied from various points
of view (see the related discussion and references in~\cite{KPV2014}).  We note that neither the classical Kawahara equation nor its generalization adduced below are integrable by the inverse scattering
transform method~\cite{MikhailovShabatSokolov1991}.

Generalized constant coefficient models related to the Kawahara equation have appeared later. For example,  long waves in a shallow liquid under ice
cover in the presence of tension or
compression were described in~\cite{Marchenko1988,Tkachenko&Yakovlev1999} by the equation $$u_t+u_x+\alpha u u_x +\beta u_{xxx}+\sigma u_{xxxxx}=0.$$
This equation is similar to the classical Kawahara equation with respect to the simple changes of variables: $\tilde x=x-t,$ where $t$ and $u$ are
not transformed, or $\tilde u=1+\alpha u$, where $t$ and $x$ are not transformed. So, all results on symmetries, conservation laws and exact solutions for this equation can be easily
derived from the analogous results for the classical Kawahara equation.

 Generalized and formal symmetries as well as local conservation laws of the constant coefficient Kawahara equations with arbitrary nonlinearity
$$
u_t+f(u) u_x +\beta u_{xxx}+\sigma u_{xxxxx}=0,\quad f_u\beta\sigma\neq0,
$$
were classified recently in~\cite{Vasicek}.

In the last years special attention is paid to  variable coefficient models of the Kawahara type. This is due to the fact that variable coefficient equations model certain real-world phenomena with more accuracy than their constant coefficient counterparts.
To the best of our knowledge, Lie symmetries of variable coefficient Kawahara equations were  studied for the first time in~\cite{Kaur&Gupta}.
Namely, the equations
$$
u_t+\alpha(t)u^n u_x +\beta(t) u_{xxx}+\sigma(t) u_{xxxxx}=0,\quad n\alpha\beta\sigma\neq0,
$$
for $n=1$ and $n=2$ were considered therein.  The presence of four arbitrary elements, $n,$ $\alpha(t),$ $\beta(t),$ and $\sigma(t)$, made
the problem of classifying Lie symmetries too difficult to get complete results without  usage of equivalence transformations.
Due to this reason a limited progress was made in~\cite{Kaur&Gupta}.
It was shown in~\cite{KPV2014} that right choice of gauging the arbitrary elements by equivalence transformations is a cornerstone for the complete solution of the problem. As a result, the extended group analysis of
the above equations for arbitrary $n\neq0$ was performed in~\cite{KPV2014}, in particular, admissible transformations, Lie symmetries and Lie reductions
were classified exhaustively as well as some solutions and local conservation laws were found therein.

Lie symmetries and local conservation laws of generalized Kawahara equations with arbitrary nonlinearity  and time-dependent coefficients,
\begin{eqnarray}\label{eq_ggKawahara}
u_t+\alpha(t)f(u)u_x+\beta(t)u_{xxx}+\sigma(t)u_{xxxxx}=0,\quad f_u\alpha\beta\sigma\neq0,
\end{eqnarray}
where $f$, $\alpha$, $\beta$ and $\sigma$ being smooth nonvanishing functions of their variables,
were studied recently in~\cite{GandariasRosaRecioAnco}.
The gauging $\sigma=1$ was performed therein to simplify the problem, however the complete classification of Lie symmetries
was not derived, in particular, because the  gauging $\sigma=1$ is not optimal.
It was shown in~\cite{VKS2016} that knowledge of normalization properties of a class of PDEs allows one to choose the optimal gauging
not by guessing but algorithmically. That is why we begin our investigation with the study of
the equivalence groupoid and normalization properties of the class~(\ref{eq_ggKawahara}) (Section~2). Then the optimal gauging of the arbitrary elements of the class is performed (Section~3). Finally,
 the complete Lie symmetry classification of such equations is carried out (Section~4).

\section{Equivalence groupoid}

It is widely known that there is no general theory for integration of
nonlinear partial differential equations (PDEs). Nevertheless, many
special cases of complete integration or finding particular exact solutions are
related to appropriate changes of variables. The transformation methods, which include in particular Lie symmetry method,
are among the most powerful analytical tools currently available for the study of nonlinear PDEs.

The systematic study of transformation  properties of classes of
nonlinear PDEs was initiated in 1991 by J.\,G. Kingston and C. Sophocleous~\cite{Kingston&Sophocleous1998}.
These authors later named the transformations related two particular
equations in a class of PDEs {\it form-preserving transformations}, because
such transformations preserve the form of the equation in a class and
change only its arbitrary elements.
Only a year later in 1992 J.\,P. Gazeau and P. Winternitz also began to
investigate such transformations in classes of PDEs calling them
{\it allowed transformations}~\cite{Winternitz92}.
Rigorous definitions and developed theory on the subject was
proposed later by R.\,O. Popovych~\cite{popo2010a}.
As formalization of notion of form-preserving (allowed)
transformations  the term {\it admissible transformations} was proposed.
In brief, an admissible transformation is a triple consisting of two
fixed equations from a given class and a nondegenerate transformation that links these
equations.
The set of admissible transformations considered with the standard
operation of composition of transformations is also called
{\it equivalence groupoid}~\cite{Popovych&Bihlo2012}.  Equivalence groupoids can be used not only in group classification problems but also in other problems related to
classes of PDEs like, for example, finding exact solutions and conservation laws,
the study of integrability~\cite{Popovych&Vaneeva2010,VPS2013}.

Equivalence transformations, which are invaluable tools of group analysis of differential equations,
generate a subset in  a set of admissible transformations.    It is important that admissible transformations are not
necessarily related to a group structure, but equivalence transformations always form a~group. An equivalence transformation  applied to any
equation from the class always maps it to another equation from the same class, while
 an admissible transformation may exist only for a specific pair of equations from the class under
consideration.

By L.\,V. Ovsiannikov, the equivalence group consists of the nondegenerate point transformations
of the independent and dependent variables and of the arbitrary elements of the class,
where transformations for independent and dependent variables are projectable on the space of these variables~\cite{Ovsiannikov1982}. After
appearance of other kinds of equivalence group the one with properties described above is called now {\it usual equivalence group}.
If the transformations for independent and/or dependent variables involve  arbitrary
elements, then the corresponding equivalence group is called   {\it generalized equivalence group}~\cite{Meleshko1994}.
If new arbitrary elements appear to depend on old ones in a nonlocal way (e.g., new arbitrary elements are expressed via integrals of old ones),
then the corresponding equivalence group is called   {\it extended equivalence group}~\cite{mogran}.
Simultaneously weakening the conditions of locality and projectability, leads to the notion of
  {\it extended generalized equivalence group}.
The notion of  {\it effective generalized equivalence group} was proposed recently in~\cite{Opanasenko2017}, this is a minimal
subgroup of the entire generalized equivalence group of a given class of PDEs
which generates the same equivalence subgroupoid of the class as the entire group does.

If any admissible transformation in a given class is induced by a transformation from its usual equivalence group, then this class is called {\it normalized}  in the usual sense. In analogous way, the notions of normalization of a class in generalized, extended and extended generalized senses are formulated~\cite{popo2010a}.
If a given class is normalized in the generalized sense, then its effective generalized equivalence group generates the equivalence groupoid of the class.

Once it is proved that certain class is normalized,  finding  the
equivalence groupoids of its subclasses becomes essentially
simpler since they are always subgroupoids of the equivalence groupoid of the initial class.
It was proved in~\cite{VPS2013} that the class
\begin{equation}\label{class_gen}
u_t=F(t)u_n+G(t,x,u,u_1,\dots,u_{n-1}),
\end{equation}
where $F\ne0,$ $G_{u_iu_{n-1}}=0,$ $i=1,\dots,n-1,$ $n\geqslant 2$,
$
u_{n}=\frac{\partial^n u}{\partial x^n},
$
$F$ and $G$ are arbitrary smooth functions of their variables, is normalized in the usual sense.  The transformation components for the variables $t,$
$x$ and $u$ of admissible transformations for the class~(\ref{class_gen}) are of the form
\begin{eqnarray}\label{EqEquivtransOfvcKdVlikeSuperclass}
\tilde t=T(t),\quad \tilde x=X^1(t)x+X^0(t),\quad \tilde u=U^1(t,x)u+U^0(t,x),
\end{eqnarray}
where $T$, $X^1$, $X^0$, $U^1$ and $U^0$ are arbitrary smooth functions of their variables with $T_tX^1U^1\neq0$.

As class~(\ref{eq_ggKawahara}) is a subclass of class~(\ref{class_gen}) with $n=5,$ transformation components for the variables $t,$ $x,$ and $u$ of admissible transformations in~(\ref{eq_ggKawahara}) can be sought  in the form~(\ref{EqEquivtransOfvcKdVlikeSuperclass}).
Following  the direct method~\cite{Kingston&Sophocleous1998}, we
suppose that equation~(\ref{eq_ggKawahara}) is similar  to an equation from the same class,
\begin{equation}\label{eq_ggKawahara_tilda}
\tilde u_{\tilde t}+\tilde\alpha(\tilde t)\tilde f( \tilde u)\tilde u_{\tilde x}+\tilde\beta(\tilde t) \tilde u_{\tilde x\tilde x\tilde
x}+\tilde\sigma(\tilde t) \tilde u_{\tilde x\tilde x\tilde x\tilde x\tilde x}=0,
\end{equation}
and these two equations are connected by a nondegenerate point~transformation of the form~(\ref{EqEquivtransOfvcKdVlikeSuperclass}).
Rewriting~(\ref{eq_ggKawahara_tilda}) in terms of the untilded variables,
we further substitute $u_t=-\alpha(t)f(u)u_x-\beta(t)u_{xxx}-\sigma(t)u_{xxxxx}$ to the derived equation.
Splitting the obtained identity with respect to the derivatives of $u$ leads to the determining
equations on the functions~$T$, $X^1$, $X^0$, $U^1$ and $U^0$, which result in the system
\begin{eqnarray}\label{ddeteq1}
&U^1_x=0,\quad \tilde\beta T_t-\beta(X^1)^3=0,\quad \tilde\sigma T_t-\sigma(X^1)^5=0,\\\label{ddeteq2}
&U^0_x\alpha f+U^1_tu+\sigma U^0_{xxxxx}+\beta U^0_{xxx}+U^0_t=0,\\\label{ddeteq3}
&\tilde\alpha\tilde f T_t=\alpha f X^1+X^1_tx+X^0_t.
\end{eqnarray}

At first we find usual equivalence group of the class~(\ref{eq_ggKawahara}). The following assertion is true.
\begin{theorem}
The usual equivalence group of the class~(\ref{eq_ggKawahara}) consists of the transformations
\begin{eqnarray*}
&\tilde t=T(t),\quad \tilde x=\delta_1x +\delta_2,\quad
\tilde u=\delta_3 u+\delta_4, \\
&\tilde f=\delta_0 f,\quad  \tilde\alpha=\dfrac{\delta_1}{\delta_0
T_t}\alpha, \quad
 \tilde\beta=\dfrac{\delta_1{}^3}{T_t}\beta,\quad  \tilde\sigma=\dfrac{\delta_1{}^5}{T_t}\sigma,
\end{eqnarray*}
where  $\delta_j,$ $j=0,1,\dots,4,$  are arbitrary constants with
$\delta_0\delta_1\delta_3\not=0$, and $T(t)$ is an arbitrary smooth function with $T_t\neq0.$
\end{theorem}

We proceed with the classification of all admissible transformations in the class~(\ref{eq_ggKawahara}).
Equation~(\ref{ddeteq2}) implies that such transformations essentially  differ for the cases $f_{uu}\neq0$ and $f_{uu}=0$.

{\bf I.} If $f(u)$ is a nonlinear function, then equations (\ref{ddeteq2}), (\ref{ddeteq3}) imply the conditions
\begin{equation*}
U^0_x=U^0_t=U^1_t=X^1_t=0, \quad \tilde\alpha\tilde f T_t=\alpha f X^1+X^0_t.
\end{equation*}
Solving these equations together with equations (\ref{ddeteq1}) we find that $X^0_t$ is proportional to the arbitrary element $\alpha,$ so the explicit form of $x$-component of admissible transformations for the subclass of the class~(\ref{eq_ggKawahara}) singled out by the condition $f_{uu}\neq0$ will be nonlocal with respect to the arbitrary element $\alpha$. To keep  equivalence transformations to be point and well defined, we extend the tuple of arbitrary elements $(f,\alpha,\beta,\sigma)$  with the additional arbitrary element $A$, that satisfies the auxiliary condition $A_t=\alpha.$
We derive the following statement.

\begin{theorem}
The subclass of the class~(\ref{eq_ggKawahara}) singled out by the constraint \mbox{$f_{uu}\neq0$}  is normalized in the extended generalized sense. Its extended generalized equivalence group~$\hat G^{\sim}$ consists of the transformations
\begin{eqnarray*}
&\tilde t=T(t),\quad \tilde x=\delta_1(x+\delta_2 A) +\delta_3,\quad
\tilde u=\delta_4 u+\delta_5, \\
&\tilde f=\delta_0 \left(f+\delta_2\right),\quad  \tilde\alpha=\dfrac{\delta_1}{\delta_0
T_t}\alpha, \quad \tilde A=\dfrac{\delta_1}{\delta_0}A+\varepsilon_0,\quad
 \tilde\beta=\dfrac{\delta_1{}^3}{T_t}\beta,\quad  \tilde\sigma=\dfrac{\delta_1{}^5}{T_t}\sigma,
\end{eqnarray*}
where  $\varepsilon_0$ and $\delta_j,$ $j=0,1,\dots,5,$  are arbitrary constants with
$\delta_0\delta_1\delta_4\not=0$, and $T(t)$ is an arbitrary smooth function with $T_t\neq0.$
The additional arbitrary element $A$ sa\-tisfies the auxiliary equation  $A_t=\alpha$.
\end{theorem}
\begin{remark}
 The group~$\hat G^{\sim}$ is also the extended generalized  equivalence group for the entire class~(\ref{eq_ggKawahara}), which is not normalized in contrast to its subclass with $f_{uu}\neq0.$ The effective generalized equivalence group~$\hat G^{\sim}$  for reparameterized class~(\ref{eq_ggKawahara}) with the extended tuple of arbitrary elements $(f,\alpha,\beta,\sigma,A)$ is a nontrivial extended generalized equivalence group
for the initial class~(\ref{eq_ggKawahara}). Indeed, this group induces the maximal subgroupoid in the equivalence groupoid of the  class~(\ref{eq_ggKawahara}) among all the equivalence groups of possible reparameterizations of this class. Moreover, the $x$-components of transformations from~$\hat G^{\sim}$ depend on the new arbitrary element $A$.
\end{remark}

{\bf II.} If $f(u)=au+b,$ where $a$ is a nonzero constant and $b$ is an arbitrary constant, then we can set $a=1$ by a~simple gauging transformation of the arbitrary elements, namely,
the transformation $\tilde f =\delta_0 f,$ $\tilde\alpha=\alpha/\delta_0$ with $\delta_0$ being a nonzero constant.
Without loss of generality we further consider the class
\begin{equation}\label{eq_Kawahara_f=u}
u_t+\alpha(t)(u+b)u_x+\beta(t)u_{xxx}+\sigma(t)u_{xxxxx}=0,
\end{equation}
where the real constant $b$ and the nonvanishing smooth functions $\alpha(t),$ $\beta(t)$ and $\sigma(t)$ are arbitrary elements.
In this case splitting of equations (\ref{ddeteq2}) and~(\ref{ddeteq3}) with respect to $u$ results in the following conditions
\begin{eqnarray*}
 & U^0_x\alpha +U^1_t=0,\quad U^0_x\alpha b +U^0_t=0,\quad \tilde\alpha U^1T_t=\alpha X^1,\\
 & \tilde\alpha T_t(U^0+\tilde b)=\alpha b X^1+X^1_tx+X^0_t.
\end{eqnarray*}
Solving these equations together with equations (\ref{ddeteq1}) we get the statement.
\begin{theorem}
The class~(\ref{eq_Kawahara_f=u})
is normalized in the extended generalized sense.
Its extended generalized equivalence group~$\hat G_1^{\sim}$ is constituted by the transformations of the form
\begin{eqnarray*}&\tilde t=T(t),\quad
\tilde x=\dfrac{\varepsilon_2x+\varepsilon_1A+\varepsilon_0}{\delta_2A+\delta_1},\quad
\tilde u=\dfrac{\varepsilon_2}{\Delta}\left((\delta_2A+\delta_1)u-\delta_2x+\delta_2bA+\varepsilon_3\right),\\&
\tilde A=\dfrac{\delta_2'A+\delta_1'}{\delta_2A+\delta_1},\quad\tilde\alpha=\dfrac{\Delta}{T_t(\delta_2A+\delta_1)^2}\alpha, \quad
\tilde \beta=\dfrac{\varepsilon_2{}^3}{T_t(\delta_2A+\delta_1)^3}\beta, \\
&
\tilde \sigma=\dfrac{\varepsilon_2{}^5}{T_t(\delta_2A+\delta_1)^5}\sigma,\quad \tilde b=\dfrac1{\Delta}\left({b\delta_1\varepsilon_2+\delta_1\varepsilon_1-\delta_2\varepsilon_0-\varepsilon_3\varepsilon_2}\right),
\end{eqnarray*}
where
 $\delta_j, \delta_j'$ $j=1,2,$ and $\varepsilon_i$, $i=0,1,2,3,$ are arbitrary constants defined up to a nonzero multiplier, with $\Delta=\delta_2'\delta_1-\delta_1'\delta_2\neq0$ and
$\varepsilon_2\not=0$. The additional arbitrary element $A$ sa\-tisfies the auxiliary equation  $A_t=\alpha$.
\end{theorem}

The equivalence groupoid of the subclass of the class~(\ref{eq_ggKawahara}) singled out by the condition $f_{uu}\neq0$ (resp. of the class~(\ref{eq_Kawahara_f=u})) is generated
by its extended generalized equivalence group~$\hat G^{\sim}$ (resp. $\hat G_1^{\sim}$).
Therefore, the class~(\ref{eq_ggKawahara}) is not normalized but it can be partitioned into two disjoint normalized subclasses each of which is normalized. Note that we have found only  effective generalized equivalence groups of the corresponding reparameterized classes. The question whether these groups  are entire generalized equivalence groups of these classes needs an additional study.

Using the results of Theorems~1 and 2 we derive the criterion of reducibility of variable coefficient generalized Kawahara equations to their constant coefficient counterparts. The following statement is true.
\begin{proposition}
An equation from the class~(\ref{eq_ggKawahara}) with variable coefficients $\alpha,$ $\beta$ and $\sigma$  is reducible to a constant coefficient equation from the
same class  if and only if the coefficients satisfy the conditions
\begin{eqnarray*}\label{criterion1}
\left(\frac\beta\alpha\right)_t=\left(\frac{\sigma\vphantom{\beta}}\alpha\right)_t=0, &\quad\mbox{for}\quad f_{uu}\ne0,&\\[1ex]
\label{criterion2}
\left(\frac1{\alpha}\left(\frac\beta\alpha\right)_t\right)_t=0,\quad
\left(\frac{\sigma\vphantom{\beta}\alpha^2}{\beta^3}\right)_t=0,&\quad\mbox{for}\quad f_{uu}=0.&
\end{eqnarray*}
\end{proposition}

\section{Gauging of arbitrary elements}
The presence of the arbitrary function $T(t)$ in the equivalence
transformations from the group $\hat G^\sim$  of the entire class~(\ref{eq_ggKawahara}) allows one to gauge
either $\alpha$ or $\beta$ or $\sigma$ to
a simple constant value, e.g., to 1.
An important question is which one of the three potential gaugings is the optimal one.
It was shown in~\cite{KPV2014,VKS2016} how to choose the optimal gauging using the normalization property of the class under consideration.
The classes  normalized in the usual sense are most convenient for investigation and the classes  normalized in the extended generalized sense is most complicated among normalized classes. If the class is not normalized one should look for the possibility of a partition of such class into normalized subclasses. In our case the class~(\ref{eq_ggKawahara}) that is not normalized can be partition into two normalized subclasses singled out by the conditions
$f_{uu}\neq0$  and $f_{uu}=0$.  Gauging the arbitrary elements will be performed separately in each subclass.

{\bf I.} If $f_{uu}\neq0,$ then Theorem~1 implies that the
subclasses of the  class~(\ref{eq_ggKawahara})  singled out by the conditions  $\beta=1$ or $\sigma=1$ will stay normalized only in the extended generalized sense, since equivalence transformations
for such subclasses will still involve $A$.
After gauging $\alpha=1$ in the subclass of~(\ref{eq_ggKawahara}) with $f_{uu}\neq0$  we obtain the class normalized in the usual sense, as in this case up to gauging equivalence transformations we can set  $A=t$  and this nonlocality disappears.

The gauging $\alpha=1$ is realized by the family of point transformations from the equivalence group $\hat G^\sim$ with
$
\tilde t=\int^t_{t_0}\!\alpha(y)\,{\rm d}y,$ $\tilde x=x,$ $\tilde u=u,$
that maps equations from the
class~(\ref{eq_ggKawahara}) with $f_{uu}\neq0$  to equations from the same class  with $\tilde
\alpha=1$, $\tilde\beta=\beta/\alpha$ and $\tilde\sigma=\sigma/\alpha$.
All results on symmetries, conservation laws, classical solutions and other related objects for equations~(\ref{eq_ggKawahara}) with \mbox{$f_{uu}\neq0$} can be found
using the similar results derived for equations from its subclass
\begin{eqnarray}\label{eq_gKawahara}
u_t+f(u)u_x+\beta(t)u_{xxx}+\sigma(t)u_{xxxxx}=0,\quad f_{uu}\beta\sigma\neq0.
\end{eqnarray}
We derive the  equivalence groupoid of
the class~(\ref{eq_gKawahara}) and formulate the following statement.
\begin{theorem}
The class~(\ref{eq_gKawahara})  is normalized in the usual sense. Its usual equivalence group~$G^{\sim}$ consists of the transformations
\begin{eqnarray*}
&\tilde t=\delta_1t+\delta_2,\quad \tilde x=\delta_3x+\delta_4 t +\delta_5,\quad
\tilde u=\delta_6 u+\delta_7, \\
&\tilde f= \dfrac1{\delta_1} \left(\delta_3f+\delta_4\right),\quad
 \tilde\beta=\dfrac{\delta_3{}^3}{\delta_1}\beta,\quad  \tilde\sigma=\dfrac{\delta_3{}^5}{\delta_1}\sigma,
\end{eqnarray*}
where  $\delta_j,$ $j=1,\dots,7,$  are arbitrary constants with
$\delta_1\delta_3\delta_6\not=0$.
\end{theorem}

{\bf II.} If $f_{uu}=0,$ then Theorem~2 implies that after the gauging $\alpha=1$ the respective subclass of the class~(\ref{eq_Kawahara_f=u}) will be normalized in the generalized sense, since the dependence on the constant arbitrary element $b$ will remain in the transformations.
So, the optimal gauging in this case is the simultaneous gauging of two arbitrary elements, $\alpha=1$ and $b=0.$
This gauging  is realized by the family of point transformations from the equivalence group $\hat G^\sim_1$ with
$
\tilde t=\int^t_{t_0}\!\alpha(y)\,{\rm d}y,$ $\tilde x=x,$ $\tilde u=u+b,$
that maps equations from the
class~(\ref{eq_Kawahara_f=u})  to equations from the same class with $\hat
\alpha=1$, $\tilde b=0$, $\tilde\beta=\beta/\alpha$ and $\tilde\sigma=\sigma/\alpha$.
Without loss of generality we can restrict ourselves with the investigation of the following class
\begin{eqnarray}\label{eq_gKawahara_u}
u_t+uu_x+\beta(t)u_{xxx}+\sigma(t)u_{xxxxx}=0,
\end{eqnarray}
instead of its superclass~(\ref{eq_Kawahara_f=u}).

The study of the equivalence groupoid of the class~(\ref{eq_gKawahara_u}) results in the following assertion.
\begin{theorem}
The class~(\ref{eq_gKawahara_u})   is normalized in the usual sense. Its usual equivalence group~$G^{\sim}_1$ consists of the transformations of the form
\begin{eqnarray*}
&\tilde t=\dfrac{\delta_4t+\delta_3}{\delta_2t+\delta_1},\  \tilde x=\dfrac{\varepsilon_2x+\varepsilon_1t+\varepsilon_0}{\delta_2t+\delta_1},
\  \tilde u=\dfrac{\varepsilon_2(\delta_2t+\delta_1)u-\varepsilon_2\delta_2x+\varepsilon_1\delta_1-\varepsilon_0\delta_2}{\Delta}, \\[0.5ex]
&\tilde\beta=\dfrac{\varepsilon_2{}^3}{(\delta_2t+\delta_1)\Delta}\beta,\quad  \tilde\sigma=\dfrac{\varepsilon_2{}^5}{(\delta_2t+\delta_1)^3\Delta}\sigma,
\end{eqnarray*}
where   $\varepsilon_i,$ $i=0,1,2,$  and $\delta_j,$ $j=1,\dots,4,$  are arbitrary constants defined up to a nonzero multiplier;
$\varepsilon_2\not=0$ and $\Delta=\delta_1\delta_4-\delta_2\delta_3\neq0.$
\end{theorem}

In the next section  we demonstrate that  the chosen gaugings allow us to solve exhaustively the group classification problems for both derived normalized subclasses of the class~(\ref{eq_ggKawahara}).

\section{Group classification}
 As in previous section we consider separately two normalized subclasses of the class~(\ref{eq_ggKawahara}), that are singled out by the conditions
 $f_{uu}\neq0$ and $f_{uu}=0.$

{\bf I.}
The group classification problem for the class~(\ref{eq_ggKawahara}) with $f_{uu}\neq0$ up to $\hat G^{\sim}$-equivalence reduces to
the similar problem for the class~(\ref{eq_gKawahara}) up to $G^{\sim}$-equivalence.
The group classification of the class~(\ref{eq_gKawahara}) is performed
using the  classical algorithm based on direct integration of the determining equations implied by the infinitesimal invariance
criterion~\cite{Olver1986,Ovsiannikov1982}.
The symmetry generators of the form $Q=\tau(t,x,u)\partial_t+\xi(t,x,u)\partial_x+\eta(t,x,u)\partial_u$
must satisfy the criterion of infinitesimal invariance,
\begin{equation*}\label{c2}
Q^{(5)}\{u_t+f(u)u_x+\beta(t)u_{xxx}+\sigma(t)u_{xxxxx}\}\Big|_{u_t=f(u)u_x+\beta(t)u_{xxx}+\sigma(t)u_{xxxxx}}=0,
\end{equation*} where  $Q^{(5)}$ is the fifth prolongation of the  vector field~$Q$~\cite{Olver1986,Ovsiannikov1982}.

The infinitesimal invariance criterion implies the determining equations simplest of which result in the following forms of $\tau,$ $\xi,$ and $\eta,$
\begin{equation*}
\tau=\tau(t),\quad
\xi=\xi^1(t)x+\xi^0(t), \quad
\eta=\left(2\xi^1(t)+\mu(t)\right)u+\eta^0(t,x),
\end{equation*}
where $\tau$, $\xi^1$, $\xi^0$, $\mu$ and $\eta^0$ are arbitrary smooth functions of their variables.

Then the rest of the determining equations have the form
\begin{eqnarray}\label{deteq1}
&\eta^0_xf+(2\xi^1_t+\mu_t)u+\eta^0_t+\eta^0_{xxx}\beta+\eta^0_{xxxxx}\sigma=0,\\\label{deteq2}
&\left((2\xi^1+\mu)u+\eta^0\right)f_u=(\xi^1-\tau_t)f+\xi^1_tx+\xi^0_t,\\\label{deteq3}
&\tau \sigma_t =(5\xi^1-\tau_t)\sigma,\quad \tau \beta_t =(3\xi^1-\tau_t)\beta.
\end{eqnarray}

\begin{table}[t!] \renewcommand{\arraystretch}{1.7}
\begin{center}
\textbf{Table~1.}
The group classification of the class\\[1ex] $u_t+\alpha(t) f(u)u_x+\beta(t) u_{xxx}+\sigma(t) u_{xxxxx}=0$,\quad $f_{uu}\alpha\beta\sigma\neq0$.
\\[2ex]
\begin{tabular}{|c|c|c|c|l|}
\hline
no.&$\ f(u)\ $&$\ \beta(t)\ $&$\ \sigma(t)\ $&\hfil Basis of $A^{\max}$ \\
\hline
0&$\forall$&$\forall$&$\forall$&$\ \partial_x$\\
\hline
1&$\forall$&$\lambda t^2$&$\delta t^4$&
$\ \partial_x,\,t\partial_t+x\partial_x$\\
\hline
2&$\forall$&$\lambda $&$\delta$&
$\ \partial_x,\,\partial_t$\\
\hline
3&$\ln u$&$\forall$&$\forall$&$\ \partial_x, t\partial_x+u\partial_u$\\
\hline
4&$\ln u$&$\lambda t^2$&$\delta t^4$&$\ \partial_x, t\partial_x+u\partial_u, t\partial_t+x\partial_x$\\
\hline
5&$\ln u$&$\lambda $&$\delta$&
$\ \partial_x,\, t\partial_x+u\partial_u,\,\partial_t$\\
\hline
6&$u^n$&
$\lambda t^\rho$&$\delta t^{\frac{5\rho+2}{3}}$&$\ \partial_x,\,3nt\partial_t+(\rho+1)n x\partial_x+(\rho-2) u\partial_u$\\
\hline
7&$u^n$&$\lambda e^{t}$&$\delta e^{\frac53 t}$&
$\ \partial_x,\,3n\partial_t+nx\partial_x+u\partial_u$\\
\hline
8&$e^u$&
$\lambda t^\rho$&$\ \delta t^{\frac{5\rho+2}{3}}\ $&$\ \partial_x,\,3t\partial_t+(\rho+1) x\partial_x+(\rho-2) \partial_u$\\
\hline
9&$e^u$&$\lambda e^{t}$&$\delta e^{\frac53 t}$&
$\ \partial_x,\,3\partial_t+x\partial_x+\partial_u$\\
\hline
\end{tabular}\\[2ex]
\parbox{90mm}{Here $\alpha(t)=1\bmod\, \hat G^\sim$,  $\rho$  and $n$ are arbitrary constants, $n\neq0,1$; $\delta$ and $\lambda$
are nonzero constants, $\delta=\pm1\bmod\, \hat G^\sim.$}
\end{center}
\end{table}
Below we give the sketch of the proof.
To find the kernel $A^\cap$  of the maximal Lie invariance algebras $A^{\max}$ of equations  (\ref{eq_gKawahara}) we split the determining equations with respect to the arbitrary elements and their derivatives, which
results in $\tau=\eta=0,$ $\xi=c_1,$ where $c_1$ is an arbitrary constant. Therefore,  $A^\cap$ is the one-dimensional algebra
$\langle\partial_x\rangle$ (Case 0 of Table~1).
At the next step we assume that $f$ is arbitrary and look for specifications of $\beta$ and $\sigma$ for which equations~(\ref{eq_gKawahara})
possess Lie symmetry extensions.
These are the cases $(\beta,\sigma)=(\lambda,\delta)$ and  $(\beta,\sigma)=(\lambda(pt+q)^2,\delta(pt+q)^4)$, where $\lambda,$ $\delta,$ $p$ and $q$
are constants and $\lambda\delta p\neq0.$ Using equivalence transformations from the group $G^\sim$ we  can set $\delta=\pm1,$
$p=1$ and $q=0.$ The respective
bases of the maximal Lie invariance algebras are adduced in Cases 1 and 2 of Table~1.

As $f_{uu}\neq0$, equation (\ref{deteq1}) implies  $\eta^0_x=\eta^0_t=0$, so, $\eta^0=c_0$ and $\mu=-2\xi^1+c_1,$ where $c_0$ and $c_1$ are arbitrary constants.
Then equation~(\ref{deteq2}) leads
to the condition
 $\xi^1_t=0$ and reduces to
\begin{equation}\label{deteq2a}
(c_1u+c_0)f_u=(\xi^1-\tau_t)f+\xi^0_t.
\end{equation}
The equations from the class~(\ref{eq_gKawahara})  possess  Lie symmetry extensions, when $f$ takes one of the following forms:
$f=\nu(a u+b)^n+\kappa,$ $n\neq0,1,$ $f=\nu e^{a u}+\kappa,$ and $f=\nu\ln(au+b)+\kappa,$ where $\nu$ and $a$ are nonzero constants, $\kappa$ and $b$ are arbitrary constants.
Up to the $G^\sim$-equivalence this list is exhausted by the cases 1. $f=u^n,$ $n\neq0,1$, 2. $f=e^u$ and 3. $f=\ln u.$
Each of these forms of $f$ should be substituted to equation~(\ref{deteq2a}), then the final forms of the coefficients $\tau,$ $\xi,$ and $\eta$ are found and the classifying equations~(\ref{deteq3}) give the possible forms of $\beta$ and $\sigma$ for which equations~(\ref{eq_gKawahara}) possess Lie symmetry extensions.
The detailed proof for the case $f=u^n, n\neq0,$ was given in~\cite{KPV2014}. The consideration of the other cases is analogous. Due to the luck of space we omit the detailed proof for these cases and formulate the final result in
the following statement.
\begin{theorem}
The kernel of the maximal Lie invariance algebras of equations from the class~(\ref{eq_ggKawahara})   with $f_{uu}\neq0$
coincides with the one-dimensional algebra $\langle\partial_x\rangle$.
All possible $\hat G^\sim$-inequivalent cases of extension of the maximal Lie invariance algebras
are exhausted by Cases 1--9 of Table~1.
\end{theorem}

{\bf II.}
In the previous section we have shown that the group classification problem for the class~(\ref{eq_Kawahara_f=u}) up to $\hat G_1^{\sim}$-equivalence reduces to
such a problem for the class~(\ref{eq_gKawahara_u}) up to $G_1^{\sim}$-equivalence. The group classification of the  class~(\ref{eq_gKawahara_u}) up to $G_1^{\sim}$-equivalence was carried out exhaustively in~\cite{KPV2014}. So we use the results derived therein and formulate the following statement.

\begin{table}[t!] \renewcommand{\arraystretch}{1.7}
\begin{center}
\textbf{Table~2.}
The group classification of the class\\[1ex] $u_t+\alpha(t) (u+b)u_x+\beta(t) u_{xxx}+\sigma(t) u_{xxxxx}=0$,\quad $\alpha\beta\sigma\neq0$.
\\[2ex]
\begin{tabular}{|c|c|c|l|}
\hline
no.&$\beta(t)$&$\sigma(t)$&\hfil Basis of $A^{\max}$ \\
\hline
0&$\forall $&$\forall$&
$\ \partial_x,\,t\partial_x+\partial_u$\\
\hline
1&
$\lambda t^\rho$&$\delta t^{\frac{5\rho+2}{3}}$&$\ \partial_x,\,t\partial_x+\partial_u,\,3t\partial_t+(\rho+1) x\partial_x+(\rho-2) u\partial_u$\\
\hline
2&$\lambda e^{t}$&$\delta e^{\frac53 t}$&
$\ \partial_x,\,t\partial_x+\partial_u,\,3\partial_t+x\partial_x+u\partial_u$\\
\hline
3&$\lambda $&$\delta$&
$\ \partial_x,\,t\partial_x+\partial_u,\,\partial_t$\\
\hline
4&\ \rule[-4ex]{0ex}{9ex}$ \dfrac{\lambda{(t^2+1)^\frac12}}{e^{3\nu\arctan t}}\ $&$\ \dfrac{\delta {(t^2+1)^\frac32}}{ e^{5\nu\arctan t}}\ $&
\ \parbox{60mm}{$\partial_x,\,t\partial_x+\partial_u,$\\[1ex] $(t^2+1)\partial_t+(t-\nu)x\partial_x+(x-(t+\nu)u)\partial_u$}\\
\hline
\end{tabular}
\\[2ex]
\parbox{110mm}{Here $\alpha=1\bmod\, \hat G^\sim_1$, $b=0\bmod\, \hat G^\sim_1$, $\rho$ and $\nu$ are arbitrary constants, $\rho\geqslant1/2\bmod \hat G_1^\sim$, $\nu\leqslant0\bmod \hat G_1^\sim$; $\delta$ and $\lambda$
are nonzero constants, $\delta=\pm1\bmod\, \hat G_1^\sim.$}
\end{center}
\end{table}

 \begin{theorem}
The kernel of the maximal Lie invariance algebras of equations from the class~(\ref{eq_Kawahara_f=u})
coincides with the two-dimensional algebra $\langle\partial_x,\,t\partial_x+\partial_u\rangle$.
All possible $\hat G_1^\sim$-inequivalent cases of extension of the maximal Lie invariance algebras are exhausted
by Cases~1--4 of Table~2.
\end{theorem}

The presented group classification reveals equations of the form~(\ref{eq_ggKawahara}) that are of more potential interest for applications and  for which the
classical  Lie reduction method can be  used. The complete result was achieved using the knowledge of transformation properties of the class namely due to partition of the class into two normalized subclasses and choosing the optimal gauging for each of these subclasses.

\begin{acknowledgement}
 O.V. would like to thank all the Organizing Committee of LT-13  and especially Prof. Vladimir Dobrev for the hospitality and support.
 The authors are grateful to Prof. Roman Popovych for invaluable discussions on the topic and also to the referee and the editor for their suggestions on the improvement of the manuscript.
\end{acknowledgement}

\end{document}